%Paper: hep-th/9304085
%From: giorgio@lpthe.jussieu.fr (Giorgio GIAVARINI)
%Date: Tue, 20 Apr 93 14:54:17 +0200

\input phyzzx
\Pubnum{}
\def\eqnalign#1{\eqname{#1}#1}
%%%%%%%%%%%%%%%%%%%%%%%%%%%%%%%%%%%%%%%%%%%%%%%%%%%%%%%%%%%%%%%%
%%Macros
%%%%%%%%%%%%%%%%%%%%%%%%%%%%%%%%%%%%%%%%%%%%%%%%%%%%%%%%%%%%%%%%
\def\ov{\overline}                          % overline
                      % dm
\def\ep{\varepsilon}                        % epsilon
                          % theta
                % cotanh
                  % acosh
                            % vector x
                            % vector y
                            % vector q
                            % vector p
\def\abs#1{\left| #1\right|}                % abs
\def\set#1{\left\{ #1 \right\} }            % set
   % Vacuum Expect. Value
\def\frac#1#2{{\textstyle{
 #1 \over #2 }}}                            % fraction
   % blank spaces

                            % F
\def\G{{\cal G}}                            % G
                            % H
                            % I
                            % C
                            % U
\def\P{{\cal P}}                            % P
                            % O
                            % D
                            % Z
                            % V
\def\L{{\cal L}}                            % L
\def\A{{\cal A}}                            % A
\def\Tr{\mathop{\rm Tr}}                    % traccia anche sul colore
                  % argomento di un numero complesso
                   % slash operator
                     % real numbers
\def\NN{{\rm I  \! N }}                     % natural numbers
                       % integers
                    % complex
\def\1{{\rm 1 \!\!\, l}}                        % 1I
%\def\d{{\rm d}}                             % derivata esterna
          % derivata totale
\def\comm#1#2{\left[ #1 \, , \, #2 \right]} % commutator
 % anti-commutator
 % scal (,)
\def\bra#1{\left\langle #1\right|}               % bra
\def\ket#1{\left| #1\right\rangle}               % ket
                         % determinant
%
% derivate parziali: accetta anche \ e {} come argomento

%
%
% prodotto scalare tra un bra ed un ket

                                 % daga
%Caligraphic characters
%
\def\A{{\cal A}}
\def\B{{\cal B}}

\def\G{{\cal G}}

%
%Greek letters
%
\def\mn{{\mu\nu}}
\def\a{\alpha}
\def\b{\beta}
\def\d{\delta}
\def\e{\eta}
\def\g{\gamma}
\def\m{\mu}
\def\n{\nu}
\def\r{\rho}
\def\s{\sigma}
\def\l{\lambda}
\def\L{\Lambda}

\def\t{\theta}

\def\ep{\varepsilon}

%
%Miscellaneous
%
\def\NN{\scriptscriptstyle N}
\def\LL{\scriptscriptstyle L}
\def\tr{\tau(\theta)}        % transfer matrix
\def\sh{\sinh}
\def\ch{\cosh}
\def\ov{\overline}
\def\sn{{\rm sn}}
\def\qh{{\hat q}}
\def\ug{\mathop{=}}
\let\ss=\scriptscriptstyle

%%%%%%%%%%%%%% FONTS
% Gothic
\font\twelvegoth=eufm10 at 12pt
\newfam\gothfam
\textfont\gothfam=\twelvegoth

% Blackboard Bold
\font\twelveBB=msym10 at 12pt
\newfam\BBfam
\textfont\BBfam=\twelveBB
\def\BB{\fam\BBfam\twelveBB}

%%%%%%%%%%%%%%%%% BEGINNING OF THE PAPER %%%%%%%%%%%%%%%%%

%\pagenumber=2
\mathsurround=2pt
\Pubnum={$\rm LPTHE~ 93/16 $}

\date{March 1993}

\titlepage

\title{{ { \seventeenrm
Exact Solution of the general \break Non Intersecting String Model}}}
\author{H.J. de Vega and G. Giavarini\foot{\rm Address after 1 May 1993:
Dipartimento di Fisica, Universit\`a di Parma
          and INFN Gruppo Collegato di Parma,
Viale delle Scienze, I-43100 Parma, ITALIA. }}
\address{{\it Laboratoire de Physique Th\'eorique et Hautes Energies,
\foot{{\rm Laboratoire Associ\'e au CNRS UA 280}}
Paris \foot{{\rm mail address: L.P.T.H.E., Tour 16 $1^{\rm er}$ \'etage,
Universit\'e Paris VI, 4 Place Jussieu F-75252, Paris Cedex 05,
FRANCE} }
}}

\vskip 1 true cm

%Abstract

\noindent We present a thorough analysis of the Non Intersecting
String (NIS) model and its exact solution.  This is an integrable
$q$-states vertex model describing configurations of non-intersecting
polygons on the lattice.  The exact eigenvalues of the transfer matrix
are found by analytic Bethe Ansatz. The Bethe Ansatz equations thus
found are shown to be equivalent to those for a mixed spin model
involving both 1/2 and infinite spin. This indicates
that the NIS model provides a representation of the quantum group
$SU(2)_{\hat q}$ ($|\hat q|\not= 1$) corresponding to spins $s=1/2$
and $s=\infty$.  The partition function and the excitations in the
thermodynamic limit are computed.

\endpage

%\doublespacing
\chapter{Introduction}

The Yang-Baxter (YB) equations assure integrability
in two-dimensional lattice models and quantum field theories. The
YB equations have the form
$$
R_{12}(\t'-\t)R_{13}(\t)R_{23}(\t')=R_{23}(\t')R_{13}(\t)R_{12}(\t'-\t)
{}~,
\eqn\YB
$$
where the $R$-matrix elements $R^{ab}_{cd}(\t)$ (with $1\leq a,b,c,d
\leq q$, $q\geq 2$) define the statistical (or Boltzman) weights for a
vertex model in two dimensions.
\REF\PerkSchultz{C.L. Schultz, Phys. Rev. Lett. {\bf 46}  (1981)
407. \nextline
J.H.H. Perk and C.L. Schultz, Physica {\bf 122A} (1983)
50; {\it Families of commuting transfer matrices in q-state vertex
models}, in: Non-linear Integrable Systems -- Classical Theory and
Quantum Theory, M. Jimbo and T. Miwa eds., World Scientific (1983).}
The general Non Intersecting String (NIS)
 model is a $q(2q-1)$ vertex model introduced
in [\PerkSchultz] as a solution of the Yang-Baxter equations.
It generalizes the $q=2$ ferroelectric
models and the $q=3$ model of Stroganov
\REF\Stroganov{Yu. G. Stroganov, Phys. Lett. {\bf A74} (1979) 116.}
[\Stroganov].
\FIG\weights{Boltzmann weights for the general NIS model. Dashed and
solid lines represent red and black indices. The remaining non-zero
weights can be obtained by crossing invariance.}

The non-zero Boltzmann weights of the model are depicted  in Fig.\weights.
We  distinguish between  indices of two different kinds that we
shall call ``red'' and ``black'', corresponding to  dashed and solid
lines respectively in the figure.  The red indices,
denoted by $\a,\b,\ldots$, take on $r$ values, whereas the black ones,
denoted by $\m,\n,\ldots$, take on the remaining $t$ values, with $r+t=q
>2$. Latin indices $a,b,\ldots$ run over all the possible values
$1,\ldots,q$.  The  values of  the  $R$-matrix elements in
Fig.\weights\ are:
$$\vbox{
\halign{# \hfil &\qquad # \hfil \cr
 $R^{\a\a}_{\a\a}(\t)=\sh\t +\sh(\e-\t)$~~,
  & $R^{\m\m}_{\m\m}(\t)=\sh\e$~~,
\cr
 &\cr
 $R^{\a\b}_{\a\b}(\t)=\sh(\e-\t)$ ~, \quad $\a\not=\b$~~,
& $R^{\m\n}_{\m\n}(\t)=e^\t
   \>\sh(\e-\t)$ ~, \quad $\m\not=\n$~~, \cr
 &\cr
 $R^{\a\m}_{\a\m}(\t)= e^{\t/2} \> \sh(\e-\t)$ ~~,
& $R^{ab}_{cd}(\t)=R^{db}_{ca}(\e -\t)$ ~~.\cr} }
\eqn\pesi
$$
The Boltzmann weights of the  NIS model, like the six-vertex ones, depend
on the spectral parameter $\t$ and  on the anisotropy parameter
 $\e$ but,
differently from the six-vertex case, $\e$ is not continuous but takes
on discrete values, solutions of the
second order equation [\PerkSchultz]
$$
(q-r-1) e^{2\e} + r e^\e -1=0 ~~.
\eqn\secondor
$$
We have then the two solutions $\e$ and ${\e'}+i\pi$ given by:
$$
e^{\e}={ \sqrt{r^2+4(q-r-1)} -r \over 2(q-r-1) } < 1
\eqn\soluno
$$
with the limiting case $e^\e=1/r$ for $q=r+1$ and
$$
e^{\e'}={ \sqrt{r^2+4(q-r-1)} +r \over 2(q-r-1) } ~~.
\eqn\soldue
$$
We see that $\e'\geq 0$ for $q\leq 2(r+1)$ and $\e'< 0$ for $q> 2(
r+1)$. It must be noticed that although the $R$-matrix becomes complex
for the solution ${\e'}+i\pi$,
the imaginary factors cancel out in the partition
function for any finite size. For simplicity,
in the following we shall consider the solution \soluno\ which corresponds
to real negative $\e$ and real $R$-matrix.

 By definition, (see
last equation in \pesi) the model is crossing invariant and $\e$ is
the crossing parameter.  As it appears clear from Fig.\weights, the
NIS model describes configurations of non-intersecting patterns of
polygons on the square lattice.  For an $N$-sites system, the spin
Hamiltonian associated to the NIS model can be written as
$$
\eqalign{
H=\sum_{j=1}^N&\left[\>  \sum_{a,b=1}^q \exp\left( {n(a,b)\over 2}\e
\right)  e^{ab}_j
\otimes e^{ab}_{j+1} +
\sum_{a,b=1}^q \left(   {n(a,b)\over 2} \sh\e-\ch\e\right)
 e^{aa}_j \otimes e^{bb}_{j+1} \right. \cr
&\quad \left. -\sh\e \sum_{a=1}^q n(a,a)\>  e^{aa}_j \otimes
e^{aa}_{j+1}  \right] ~~, }
$$
where  $n(a,b)$ is the number of black indices among $a$ and $b$
($n(a,b)=0,1$ or $2$) and
$(e^{ab})_{cd}=\delta_{ac}\delta_{bd}$.

\REF\Berg{B. Berg, M. Karowsky, P. Weisz and V. Kurak, Nucl. Phys.
{\bf B134} (1978) 125}
\REF\PerkWu{J.H.H. Perk and F.Y. Wu, J. Stat. Phys. {\bf 42} (1986) 727.}
\REF\Zamolodchikov{A.B. Zamolodchikov, Mod. Phys. Lett. {\bf A6} No.
19 (1991) 1807.}
\REF\Deguci{T. Deguchi, M. Wadati and Y. Akutsu, Adv. Stud. in Pure Math.
{\bf 19} (1989)  193.}
\REF\Wuknots{F.Y. Wu, {\it Knot theory and statistical mechanics}, preprint.}
Several papers dealing with the special  case $t=0$, also called
the ``separable vertex model'', already appeared in the literature
[\PerkSchultz-\Zamolodchikov].
In particular, $r=q=3$  corresponds to the
model considered  by Stroganov [\Stroganov].  Equivalence between the
separable NIS model with a $q^2$-Potts model has been established in
[\PerkWu].
Analytical continuation to $-2<q<2$
of the separable case corresponds to the scaling limit
of the $O(n)$ vector model with $-2<n<2$ away from the critical point
[\Zamolodchikov]. As it is well known,
the partition function of integrable vertex models
can be used for computing knot invariants [\Deguci].  The
separable NIS model leads then to the Jones polynomial [\Wuknots].

 In this paper we find the exact  solution of the
NIS model in its full generality ($t\not= 0$)
 by means of  analytic Bethe Ansatz (BA). We show that with
periodic boundary conditions on a $N$-sites lattice ($N$ even), the
eigenvalues of the transfer matrix $\tau(\t)$ of the general NIS model
have the form
$$
\L(\t,\vec \l)=
  F(\t) \>\sinh^N(\e-\t) \> e^{\t B}
+  \overline{F(\e-\t)} \> \sinh^N\t \> e^{(\e-\t) B} ~~,
$$
where  $B$ is an integer ($0\leq B\leq N$),
$\vec\l=(\l_1,\ldots,\l_m)$ and
$$
F(\t)=e^{i{\omega}}
\prod_{j=1}^m {\sin(\l_j-i\t-i\e /2) \over \sin(\l_j-i\t+i\e /2)}  ~~.
$$
The numbers $\l_j$ in $F(\t)$ are  obtained by solving
the BA equations which read, for the NIS model, (see Sect.3):
$$
\left( {\sin(\l_k+i\e /2) \over \sin(\l_k-i\e /2) } \right)^N \>
e^{2i(\l_kB-\omega)}=
 \prod_{ {i=1 \atop i\not= k} }^m
 {\sin(\l_k-\l_i+i\e) \over \sin(\l_k-\l_i-i\e) }  ~,
\qquad 1\leq k \leq m ~.
\eqn\BAEI
$$
Here $\omega$ is a real phase. We conjecture (see Secs.3 and 6 for more
details and a proof in the limit $N\to\infty$)
that $e^{i\omega}$ is in fact a root of unity  i.e.
$$
e^{i\omega l}=1
$$
for some integer $l$. The eigenvectors of $\tau(\t)$ are thus labelled
by $B$, $\omega$ and $m$. The antiferromagnetic ground state
is obtained for $B=0$, $m=N/2$ and $\omega=0$. The corresponding
eigenvalue of $\tau(\t)$ coincides {\it exactly} with the
antiferromagnetic ground state of the six-vertex model with the
identification $\g=-\e$. Notice that the BA equations \BAEI\ are
identical to those of the six-vertex model if $\g=-\e$, {\it except}
for the phase factor $e^{2i(\l_k B-\omega)}$.
\REF\Ogi{E. Ogievetski and P.B. Wiegmann, Phys. Lett. {\bf B168}
(1986) 360. \nextline
N.Yu. Reshetikhin and P.B. Wiegmann, Phys. Lett. {\bf B189}
(1987) 125.}
\REF\mixspin{H.J. de Vega and F. Woynarovich, J. Phys. {\bf A}: Math. and
Gen. {\bf 25} (1992) 4499.}
\REF\babujian{H.M. Babujian and A.M. Tsvelick, Nucl. Phys. {\bf B265}
[{\bf FS}15] (1986) 24.}

We recall that for a spin-$s$ model the L.H.S. of the BA equations
is given by [\babujian]
$$
\left( {\sin(\l_k+is\g ) \over \sin(\l_k-is\g ) } \right)^{N_s} ~~,
$$
whose  $s\to\infty$ limit is $\exp (-2i\l_k N_s )$. Moreover, as
shown in ref. [\mixspin], integrable models with mixed spins can be
constructed and solved by BA. This evidence suggested us that a deep
link  exists between the NIS model and a model containing both
spins 1/2 and infinite spins. We are in fact able to {\bf prove} that
the BA equations and the  eigenvalues of the NIS model
{\bf coincide} with those of
a model containing $N$ spins 1/2 and $N_s$ spins $s$ in the limit in
which $s$ is infinite. The proof of this equivalence
is given is Sec. 6 in the limit $N,N_s \to\infty$. As a byproduct we
find that the phases $e^{i\omega}$ are roots of unity. In this
context, the discrete parameter $B$ turns out to be the contribution
of the $\infty$-spins to the total $z$-component of the spin (normalized to
one) for each BA state.

 It is  known that it is  possible to associate a family of solutions
of the YB algebra
\YB\ to each Lie algebra $G$ [\Ogi].
Trigonometric (hyperbolic) solutions usually provide highest weight or
cyclic representations of the corresponding quantum group $G_\qh$.
Here $\qh=e^{i\g}$ ($e^\g$) where $\g$ is the anisotropy parameter in
the $R$-matrix.
As the discussion above indicates,   the NIS model should  provide a
new  representation of the quantum group $SU(2)_\qh$
with $\qh=e^\e$ (notice that $|\qh|\not= 1$). This representation is
absent in the ``classical'' case $\qh=1$ and is probably
indecomposable containing $s=1/2$ and $s=\infty$.

The plan of the paper is as follows. In Sec. 2  we summarize the
general properties of the NIS $R$-matrix and transfer matrix. In Sec.
3 we construct the BA solutions of the NIS model through analytic BA.
Sec. 4 contains  checks for large values of the spectral parameter
$\t$ of the results obtained by BA. In Sec. 5 we find the free energy
and the excitations in the thermodynamic limit. Finally, in Sec. 6,
the connection between the NIS model and a spin 1/2-spin $\infty$
model is derived.

\chapter{General properties of the NIS model}
For the general NIS model we find, by direct computation, the following
properties for the $R$ matrix
\item{a)} Regularity at $\t=0$
$$
R(0)=\sh\e \> \1 ~~.
\eqn\regularity
$$
\item{b)} Unitarity
$$
R(\t)R(-\t)=\r(\t) \> \1 ~~,
\eqn\unitarity
$$
where $\r(\t)=\sh(\e-\t)\sh(\e+\t)$
\item{c)} Quasi periodicity
$$
R(\t+i\pi)=(u\otimes 1)R(\t)(u^{-1}\otimes h) ~~,
\eqn\quasiper
$$
where $u$ and $h$ are the diagonal matrices   $u_{ab}=u_a
\d_{ab}$ and $h_{ab}= h_a\d_{ab}$ with
$$
\matrix{
 &u_\a =1~~, & u_\mu =i~~, \cr
 & h_\a =-1~~, &h_\m=1~~. }
\qquad\left\{ \matrix{1\leq\a \leq r \cr
        r+1\leq \m \leq q} \right.
$$
\item{d)} Invariances. A Yang-Baxter algebra is said to be invariant
under a transformation  group $\G$ if the $R$-matrix
satisfies
\REF\deVega{H.J. de Vega, Int. J. Mod. Phys. {\bf A4} No. 10 (1989)
2371.}
[\deVega]
$$
\comm{g\otimes g}{R(\t)}=0 ~,
\qquad \forall \t\in{\BB C}\>,\quad \forall g\in\G ~~.
$$
In the separable case, $r=q$, from eqs.\pesi\ it follows
$$
g g^T=1 ~~,
$$
that is we have invariance under the orthogonal group.

In the other
limiting case in which all the indices are of the black kind ($r=0$),
we find
$$
(gg^T)_{\m\m}=1\>, \qquad \forall \mu
$$
and
$$
e^{-(\t-\e)}\sum_{\s\not=\tau}\>g_{\m\s}g_{\n\s} +
e^{\t-\e}\>g_{\m\tau}g_{\n\tau} =0 ~, \qquad \m\not=\n
$$
which, for the linear independence of the factors $e^{\pm (\t-\e)}$
yields
$$
g_{\m\r}g_{\n\r}=0\>, \qquad \m\not=\n ~~.
$$
Thus $\G$ is formed by matrices $g$ with only one element different
from zero, equal to $\pm 1$, in each row and column.  The
symmetry group is in this case ${\cal S}_q\times \left({\BB
Z}_2\right)^q $,  the direct
product of ${\cal S}_q$, the symmetric group of $q$ elements, and
$\left( {\BB Z}_2\right)^q$.

In the general case ($r\not= 0,q$), there is no symmetry
mixing the black and red sectors of the theory. Therefore the symmetry
group is simply given by the direct product
$$
\G = O(r)\times {\cal S}_t \times \left({\BB Z}_2\right)^t ~~,
$$
where $t=q-r$ is the number of black indices.

We now turn our attention to some properties of the transfer matrix
$\tau(\t)$ which, for a $N$-sites system, is defined as
$$
\tau(\t)=\Tr_{\cal A}\left( R_{{\cal A}1}(\t)
R_{{\cal A}2}(\t) \ldots
R_{{\cal A}N}(\t) \right) ~~.
$$
{}From  property \quasiper\ of quasiperiodicity one gets for the
transfer matrix
$$
\eqalign{
\tau(\t+i\pi)=&\Tr_{\cal A}\left( u_{\cal A} R_{{\cal A}1}(\t)
u^{-1}_{\cal A} h_1
 u_{\cal A} R_{{\cal A}2}(\t) u^{-1}_{\cal A} h_2 \ldots
 u_{\cal A} R_{{\cal A}N}(\t) u^{-1}_{\cal A} h_N \right) \cr
=&\tau(\t) {\cal H} ~~, }
$$
with ${\cal H}=\bigotimes_{j=1}^N h_j$. Next we note that the matrix $h$ is
a  symmetry of $R(\t)$ (property {\it d})
and that $(hu)^{-1}=u^{-1}h$ then
$$
R_{12}(\t+i\pi)=(hu\otimes h)R_{12}(\t) (u^{-1}h\otimes 1) ~~.
$$
So that
$$
\eqalign{
\tau(\t+i\pi)=&\Tr_{\cal A}\big[ (hu)_{\cal A} h_1 R_{{\cal A}1}(\t)
(u^{-1}h)_{\cal A} \>
 (hu)_{\cal A} h_2 R_{{\cal A}2}(\t) (u^{-1}h)_{\cal A}  \ldots \cr
&\qquad \times
 (hu)_{\cal A} h_N R_{{\cal A}N}(\t) (u^{-1}h)_{\cal A}  \big] \cr
=&{\cal H} \> \tau(\t) ~~. }
$$
Thus,
$$
\tau^2(\t+i\pi)=\tau(\t){\cal H}^2\tau(\t)=\tau^2(\t)
\eqn\quadra
$$
since ${\cal H}^2=1$. This shows that configurations on the square
lattice with periodic boundary conditions  of
the NIS model are invariant under the change $\t\to\t+i\pi$.

The key property for the applicability of the analytic BA technique to
the general NIS model is the inversion relation satisfied by the
transfer matrix, namely
$$
\tau(\t+\e)\>\tau(\t)= \left[ \sh(\e+\t)\>\sh(\e-\t) \right]^N +
\sh^N\t\> \widetilde \tau(\t) ~~,
\eqn\inversion
$$
where $\widetilde \tau(\t)$ is a matrix regular at $\t=0$.
In order to prove \inversion\ it is convenient to switch to the
$S$-matrix, which is related to the $R$-matrix by
$S^{ab}_{cd}(\t)=R^{ab}_{dc}(\t)$.   We
notice that, from eq. \regularity\ and crossing invariance we have
$$
S^{ab}_{cd}(\e)=\sh\e \> \d^{a}_b \d^{c}_d ~~;
$$
so $S(\e)$ is proportional to
$$
\P={1\over q} \sum_{a,b=1}^q \ket{aa} \bra{bb}
$$
which is a  projector onto a one-dimensional subspace.
The Yang-Baxter equation for the $S$-matrix reads
$$
S_{12}(\t'-\t)S_{13}(\t)S_{23}(\t')=S_{23}(\t')S_{13}(\t)S_{12}(\t'-\t)
{}~~.
\eqn\YBE
$$
Setting $\t'-\t=\e$ we get
$$
\P_{12}S_{13}(\t)S_{23}(\e+\t)=S_{23}(\e+\t)S_{13}(\t)\P_{12}
$$
which, left multiplied by $\P^{\perp}_{12}=1-\P_{12}$,  yields
$$
\P_{12}S_{13}(\t)S_{23}(\e+\t)\P^{\perp}_{12}=0 ~~.
$$
Therefore,  the matrix $S_{13}(\t)S_{23}(\e+\t)$ takes the block
triangular form
$$
S_{13}(\t)S_{23}(\e+\t)=
\pmatrix{ A_3(\t) & 0 \cr
          \ast  & B_3(\t) \cr}
\eqn\btriang
$$
in the basis where $\P_{12}$ has the canonical form
$$
\P_{12}=\pmatrix{1 &0\cr 0&0\cr} ~~.
\eqn\canon
$$
The matrix notation in eqs. \btriang\ and \canon\ refers to the tensor
product ${\cal V}_1\times{\cal V}_2$ of the  vector spaces ${\cal
V}_1$ and ${\cal V}_2$ so
that the matrix elements are matrices in the vector space ${\cal
V}_3$. From eq. \btriang\ we can determine $A(\t)$ which is given by
$$
\P_{12}S_{13}(\t)S_{23}(\e+\t)\P_{12}=A_3(\t)\P_{12} ~~.
$$
The $S$-matrix defined by the $R$-matrix elements in eqs.\pesi\ is
time-reversal invariant
$$
S(\t)=S(\t)^T ~~.
$$
By using time reversal, unitarity [eq.\unitarity], crossing invariance
[$\>S_{12}(\t)^{T_2}=S_{12}(\e-\t)\>$] of  $S(\t)$ and from
the explict form of $\P$ we get
$$
A(\t)=\r(\t) \, \1 ~~.
$$
This shows that the matrix $A(\t)$ is diagonal.

Analogously, we can show that the matrix $B(\t)$ is zero at $\t=0$.
{}From eq.\btriang\ we have
$$
\P_{12}^\perp S_{13}(0) S_{23}(\e) \P_{12}^\perp = B_3(0)
\P_{12}^\perp ~~.
\eqn\Beq
$$
As consequence of crossing and regularity, it is
$$
S_{13}(0)S_{23}(\e)\propto P_{13}\P_{23} ~~,
$$
where $P_{13}$ is the exchange operator in the spaces 1 and 3:
$$
P_{13}=\sum_{a,b=1}^q \ket{ab}\bra{ba} ~~.
$$
It is then a matter of straightforward algebra to show that the LHS,
and therefore $B(0)$, in
\Beq\ is zero. With no loss of generality, we can then rewrite
eq. \btriang\ in  the form
$$
S_{13}(\t)S_{23}(\e+\t)=
\pmatrix{ \r(\t)  & 0 \cr
          \ast  & \sh\t \> \widetilde S_{3}(\t) \cr} ~~,
\eqn\btriang
$$
where
$\widetilde S_{3}(\t)$ is a $(q^2-1)\times (q^2-1)$ dimensional
matrix in ${\cal V}_1\times{\cal V}_2$ regular at $\t=0$ whose
elements are matrices in  ${\cal V}_3$.

Consequently
$$
\eqalign{
\tau(\t)\tau(\e+\t)=&\Tr_{{\cal A}{\cal B}}
\big(S_{\A 1}(\t)S_{\B 1}(\e+\t)S_{\A 2}(\t)S_{\B 2}(\e+\t)\ldots
S_{\A N}(\t)S_{\B N}(\e+\t) \big)  \cr
&=\r(\t)^N + \sh^N\t \>
\Tr_{{\cal A}{\cal B}}\left(\widetilde S_1(\t)\ldots\widetilde S_N(\t)
\right) ~~,}
$$
with $\rho(\t)$ given in eq. \unitarity. This proves eq. \inversion.

\chapter{Analytic Bethe Ansatz (BA) for the general NIS model}

We start by constructing a family of eigenstates for the transfer
matrix that we shall call reference states. Although these states
provide only a small portion of the spectrum of $\tr$, their
knowledge is an essential ingredient for the formulation of the
analytic BA which, in turn, will generate the whole spectrum of $\tr$.

The action of $\tr$ on a general vector $\ket{a_1,\ldots,a_N}$, is given by
$$
\tr\ket{a_1,\ldots,a_N}=\sum_{\set c ,\set b =1}^q R^{c_1 a_1}_{b_1 c_2}(\t)
R^{c_2 a_2}_{b_2 c_3}(\t) \ldots  R^{c_N a_N}_{b_N c_1}(\t)
\ket{b_1,\ldots,b_N} ~.
\eqn\actiontr
$$

If we restrict our attention to those states such that [\PerkSchultz]
$a_{i+1}\not=a_i$ for $i=1,\ldots,N$, then eq.\actiontr\ takes a
particularly simple form. Indeed, in \actiontr\ let us consider the
following two possibilities: (i) $c_1\not= a_1$ and (ii)
$c_1\not=a_N$.
In the case (i), $R^{c_1 a_1}_{b_1 c_2}(\t)$ is non-zero only if
$b_1=c_1$ and $c_2=a_1$. This implies that in the next factor $ R^{c_2
a_2}_{b_2 c_3}(\t)$, since $a_1\not= a_2$, we must have $b_2=a_1$ and
$c_3=a_2$ and so on,
%. This pattern propagates along the product of weight factors,
 hence the sum in \actiontr\
reduces to  just one term:
$$
\prod_{i=1}^N R^{a_i a_{i+1}}_{a_i  a_{i+1}}(\t)
\ket{a_N,a_1,a_2,\ldots,a_{N-1}} ~~.
$$
Here and in the following we set $a_{N+i}=a_i$.

If instead $c_1\not= a_N$, then $c_N=a_N$ and $c_1=b_N$ in
$R^{c_N a_N}_{b_N c_1}(\t)$.
In a way similar to case (i), this time
moving backwards in the product of weight factors in eq.\actiontr, we get
$$
\prod_{i=1}^N R^{a_i a_{i}}_{a_{i+1}  a_{i+1}}(\t)
\ket{a_2,\ldots,a_{N},a_1} ~~.
$$
As we have just seen,  case (ii) implies $a_1=c_1$. Thus, the two
cases considered above  exhaust all the possibilities. The total result
is then
$$
\eqalign{
\tr\ket{a_1,\ldots,a_N}= &\prod_{i=1}^N R^{a_i a_{i+1}}_{a_i  a_{i+1}}(\t)
\ket{a_N,a_1,a_2,\ldots,a_{N-1}} \cr
& \quad +\prod_{i=1}^N R^{a_{i\phantom{+1}} a_{i}}_{a_{i+1}  a_{i+1}}(\t)
\ket{a_2,\ldots,a_{N},a_1} ~~. }
\eqn\prelim
$$
Now, the value of the weights $R^{ab}_{ab}(\t)=R^{aa}_{bb}(\e-\t)$
depends on whether $a$ and $b$ are black or red indices. From
eqs.\pesi\ we find
$$
R^{ab}_{ab}(\t)=\sinh(\e-\t) \> e^{{\t \over 2} n(a,b)} ~~,
$$
where $n(a,b)$ is the number of black indices among $a$ and $b$
($n(a,b)=0,1$ or $2$). Thus, by taking into account that each index appears
twice in the products, we obtain
$$
\prod_{i=1}^N R^{a_i a_{i+1}}_{a_i  a_{i+1}}(\t) = \sinh^N(\e-\t)
\> e^{\t B},
\qquad \prod_{i=1}^N R^{a_{i\phantom{+1}} a_{i}}_{a_{i+1}  a_{i+1}}(\t) =
\sinh^N\t \> e^{(\e-\t) B}
$$
where $B$ is the number of black indices in the state
$\ket{a_1,\ldots,a_N}$.
Since the effect of applying  $\tr$ amounts, apart from an overall
factor,  to shifting the indices $a_i$ to the left or to the the
right, eigenstates of $\tr$ can be  obtained by constructing ``plane
waves''  of the form
$$
\Psi(\ep_{\LL}^j;a_1,\ldots,a_N) = \sum_{k=0}^{L-1} \ep_{\LL}^{jk}
\ket{a_{1+k},\ldots,a_{N+k}}, \qquad
\matrix{ a_{i+1}\not= a_i \cr
         a_{N+i} = a_i }
\eqn\refstate
$$
with $\ep_{\LL}=e^{{2\pi i\over L}}$. In
 eq.\refstate\ $L$ is the smallest integer such
that $\ket{a_{L+1},\ldots,a_{L+N}}=\ket{a_1,\ldots,a_N}$ and $1\leq j<L$.
Depending on the choice of the indices $a_1,\ldots a_N$, $L$ takes the
values $2\leq L \leq N$, with $N/L$ integer.
These states are eigenvectors of $\tr$ with eigenvalue
$$
\L(\ep_{\LL}^j,\t)=\ep_{\LL}^j \sinh^N(\e-\t) \> e^{\t B}
+ \ep_{\LL}^{-j} \sinh^N \t \> e^{(\e-\t) B}
\eqn\refvalue
$$
and they provide a basis in the sector of
the Hilbert space spanned by  vectors
$\ket{a_1,\ldots,a_N}$ with $a_{i+1}\not= a_i$. Indeed
$$
\ket{a_1,\ldots,a_N}={1\over L} \sum_{j=1}^L \Psi
(\ep_{\LL}^j;a_1,\ldots,a_N)
$$
owing to  $\sum_{j=1}^L \ep_{\LL}^j =0$.
Notice that the eigenvalues \refvalue\
are highly degenerate since they do not depend on the particular
choice of $\ket{a_1,\ldots,a_N}$ as long as $B$ is left unchanged.

Clearly, $B\equiv 0$ in the separable case. In the  case  $r=0$, it is
still possible to  build reference states for which the eigenvalues
are of the form \refvalue\ with $B\not= N$ by forming
suitable linear combinations of states with one or more couples of
equal indices. For example
starting from states of the form
$$
\ket{\m\m\r_3\ldots\r_N} ~~,
\eqn\couple
$$
with $\r_i\not=\r_{i+1}$ and $\m\not= \r_3,\r_N$, as before we form
the plane waves
$$
\ket{\ep_{\NN}^j;\m}=\sum_{k=0}^{N-1} \ep_{\NN}^{jk}
\ket{\r_{1+k},\ldots,\r_{N+k}} ~~,
\eqn\planecouple
$$
$$
\r_i\not=\r_{i+1} \quad {\rm if} \quad i\not= 1\>, \quad
\r_1=\r_2=\m ~, \quad \r_{N+i}=\r_i ~~.
$$
Then,  states of the kind
$$
\Psi (\ep_{\NN}^j;\m,\n) = \ket{\ep_{\NN}^j;\m} - \ket{\ep_{\NN}^j;\n}
%\eqn\lowB
$$
%$$
%\eqalign{
%\r_i\not=&\r_{i+1} \quad {\rm if} \quad i\not= 1,2\>, \quad
%\r_1=\r_2=\m ~, \quad \r_{N+i}=\r_i\cr
%\s_i\not=&\s_{i+1} \quad {\rm if} \quad i\not= 1,2\>,  \quad \s_1=\s_2=\n
%~, \quad \s_{N+i}=\s_i}
%$$
are eigenstates of the transfer matrix with
 eigenvalues of the form \refvalue\ with $B=N-2$.
By taking appropriate linear combinations of plane waves
 involving states  with  two and more couples of equal neighboring
indices,  we span all the possible even values of $B$ from $N$
to $0$ ($N$ is supposed to be even).
This also shows that, in the case where only black indices are allowed, $B$ is
bound to take on even values.

We are now in the position to formulate the analytic BA for the
general NIS model. Eqs. \inversion, \quadra\ and crossing invariance
of  $\tr$ imply the following properties
for the  eigenvalues $\L(\t)$ of $\tau(\t)$:
$$
\L(\t+\e)\>\L(\t)=\sinh^N(\e+\t)\>\sinh^N(\e-\t) + \sh^N\t \>\widetilde\L(\t)
{}~~, \eqn\propI
$$
$$
\L(\t+i\pi)^2=\L(\t)^2 ~~,
\eqn\propII
$$
$$
\ov{\L(\t)}=\L(\e-\t) ~~,
\eqn\propIII
$$
where $\widetilde\L(\t)$ is a regular function at $\t=0$.

Starting from  expression \refvalue\ of the reference eigenvalues,
we propose for the $\L(\t)$ the general analytical form
$$
\L(\t)=
  F(\t) \>\sinh^N(\e-\t) \> e^{\t B}
+  G(\t) \> \sinh^N\t \> e^{(\e-\t) B} ~~,
\eqn\ansatz
$$
where  $B$ is an integer ($0\leq B\leq N$) and $F(\t)$ and $G(\t)$
are functions to be determined. As previously discussed, $B\equiv 0$ in
the separable case and $B$ even if $r=0$.
{}From the explicit expression of the $R$-matrix elements and  eq.\propII\
it follows that
$\L(\t)$ must be a polynomial in $e^\t$ of degree at most $2N$
and of definite parity
under the change $\t\to\t+i\pi$. Therefore with no  loss of generality
we can take $F(\t)$ and
$G(\t)$ to be rational functions of $e^{2\t}$ with the same number of zeros
and poles. Moreover, in order to obtain $\L(\t)$ entire in
$e^\t$, the poles of $F(\t)$ and $G(\t)$ must have the same location,
so that the corresponding singular contributions may be made to
cancel. This last requirement is what will give us the Bethe-Ansatz equations.
Summarizing we look for $F(\t)$ and $G(\t)$ of the form
$$
\eqalignno{
F(\t)= & e^{\phi} \prod_{i=1}^m {e^{2\t} -z_i \over e^{2\t} -\xi_i}\>,
& \eqnalign\expF \cr
G(\t)= & e^{\psi} \prod_{i=1}^m {e^{2\t} -\zeta_i \over e^{2\t} -\xi_i}\>,
& \eqnalign\expG \cr
}
$$
where $\phi,\psi,z_i,\zeta_i$ and $\xi_i$ are complex parameters to be
determined. The expression \ansatz\ and the form of  $F(\t)$ and $G(\t)$
shows that the parity of $\Lambda(\t)$ under the change
$\t\to\t+i\pi$ is entirely determined by the choice of $B$.

{}From eqs.\propI, \propII\ and \propIII\ it follows
$$
F(\t)\>G(\t-\e)=1, \qquad \ov{G(\t)}=F(\e-\t)
\eqn\eqG
$$
or
$$
F(\t)\> \ov{F(-\t)}=1 ~~.
\eqn\chiave
$$
By substituting the expression \expF\ into eq. \chiave\ we obtain
$$
e^{2{\rm Re}\phi} \prod_{i=1}^m {e^{2\t} -z_i \over e^{2\t} -\xi_i}
= \prod_{i=1}^m {1-e^{2\t}\bar\xi_i  \over 1- e^{2\t}\bar z_i } ~~.
\eqn\inter
$$
Zeros and poles on both sides of eq.\inter\ coincide if $\bar\xi_i
z_i =1$.  Equating the residues at the
poles of both members of eq. \inter\ imposes
$$
e^{2 {\rm Re}\phi}=\prod_{i=1}^m \abs{\xi_i}^2 ~~.
$$
On the other hand eqs. \eqG\ give
$$
e^{\psi} \prod_{i=1}^m {e^{2\t} -\zeta_i \over e^{2\t} -\xi_i}
=
e^{-\phi}
\prod_{i=1}^m {e^{2\t-2\e} -\xi_i \over e^{2\t-2\e} -\bar\xi_i^{-1}} ~~,
$$
that is: $\psi=-\phi$, $\zeta_i=e^{2\e}\xi_i$ and $\xi_i=e^{2\e}\bar
\xi_j^{-1}$. It is  then natural to define  $\m_i=e^{-\e}\xi_i$ so that
$$
\zeta_i=e^{3\e}\m_i, \qquad
{\rm Re}(\phi) = m\e \quad
{\rm with}\quad \set{\m_i}=\set{\bar {\m}_j^{-1}} ~~.
$$
The final result is
$$
\eqalignno{
F(\t)= & e^{i{\omega}}
\prod_{i=1}^m {e^{2\t+\e} -\m_i \over e^{2\t} -e^\e\m_i}\>,
& \eqnalign\expFfin \cr
G(\t)= & e^{-i{\omega}}
\prod_{i=1}^m {e^{2\t-\e} -e^{2\e}\m_i \over e^{2\t} -e^\e\m_i}\>,
& \eqnalign\expGfin \cr
}
$$
with $\omega={\rm Im}\phi$.

Reinserting eqs.\expGfin\ and \expFfin\ back in \ansatz, and imposing
the vanishing of the residues at the poles of $\Lambda(\t)$, so to
actually obtain a polynomial in $e^\t$, we get the
Bethe-Ansatz (BA) equations. The singularities of $F(\t)$ and $G(\t)$
are at  those values $\t_k$ of $\t$ such that $e^{2\t_k}=e^\e \m_k$.
The BA equations are thus:
$$
\left( {\sh\t_k \over \sh(\e-\t_k) } \right)^N \> e^{(\e-2\t_k)B}=
e^{2i\, \omega} \prod_{ {i=1 \atop i\not= k} }^m
 {\sh(\t_k-\t_i+\e) \over \sh(\t_k-\t_i-\e) }
$$
which, after the change of variable $\t_k=\e /2 -i\l_k$, read
(assuming $N$ even)
$$
\left( {\sin(\l_k+i\e /2) \over \sin(\l_k-i\e /2) } \right)^N \>
e^{2i(\l_kB-\omega)}=
 \prod_{ {i=1 \atop i\not= k} }^m
{\sin(\l_k-\l_i+i\e) \over \sin(\l_k-\l_i-i\e) } ~~.
\eqn\BAE
$$
The eigenvalues of the transfer matrix have thus the form
$$
\eqalign{
\L(\t,\vec \l)=&
\left(\prod_{j=1}^m {\sin(\l_j-i\t-i\e /2) \over \sin(\l_j-i\t+i\e /2)
} \right)
\>\sinh^N(\e-\t) \> e^{\t B+i{\omega}} \cr
&\quad + \>
\left( \prod_{j=1}^m {\sin(\l_j-i\t+3i\e /2) \over \sin(\l_j-i\t+i\e
/2) }\right)
\> \sinh^N\t \> e^{(\e-\t) B-i{\omega} } ~~, }
\eqn\final
$$
with $\vec\l=(\l_1,\ldots,\l_m)$ solution of \BAE.
We see that the BA equations of the general NIS model
are very similar but not identical to
those  of the six-vertex model. First, in the six-vertex with periodic
boundary conditions the  phase factor $e^{i\omega}$ is absent; second and
more important, a $\l$-dependent phase factor $e^{2i\l_kB}$ never
appears in the six-vertex model.
It is usually possible to determine the phase $\omega$ by studying the
asymptotic behaviour for $\t\to\infty$ of the transfer matrix
\REF\Reshetikhin{N.Yu. Reshetikhin, Sov. Phys. JETP {\bf 57} (1983) 691.}
[\Reshetikhin]. For the
general NIS model the diagonalization of the transfer matrix in this
limit is not simpler than the diagonalization for finite values of
$\t$. An analysis of the large  $\t$ asymptotics  of $\tau(\t)$ will be
the subject of the  the next section.
In the separable case $e^{i\omega}$ is known to  be a  root of unity
[\PerkSchultz].  Computations on small size
examples and the analysis of the next section
led us to conjecture that the same result holds true
in  the general case; in Sec. 6 we provide a proof of this result in the
thermodynamic limit.

To analyze the  structure of the BA equations, we take as usual the
logarithm of eq. \BAE\ which reads ($\g=-\e>0$)
$$
N\Phi(\l_k,\g /2)+2\l_k B -2\omega
-\sum_{ {i=1 \atop i\not= k} }^m \Phi(\l_k-\l_i,\g) = 2\pi I_k
\eqn\LBAE
$$
where
$$
\Phi(\l,\g)\equiv i \log\left[ \sin(i\g+\l) \over \sin(i\g-\l) \right]
\eqn\FI
$$
and the $I_k$ are half integers.
We choose the branches of the logarithm in \FI\ in such a way that
the function $\Phi(x,\g)$ for real $x$ is continuous, odd
 and monotonically increasing for $-\pi / 2\leq \l\leq \pi
/2$. In particular
$$
\Phi\left(\pm{\pi\over 2},\g \right)=\pm\pi ~~,
\eqn\CFI
$$
$$
\Phi(\l+\pi,\g)- \Phi(\l,\g)=2\pi ~~.
\eqn\DISC
$$

As customary we introduce  the so called
counting function $Z_N(\l)$ [\deVega] that in the
present case takes the form
$$
Z_N(\l)={1\over 2\pi} \left[ \Phi(\l,\g /2)+ 2\left(b\l-{\omega\over N}\right)
-{1\over N}  \sum_{ {i=1} }^m \Phi(\l-\l_i,\g) \right]
\eqn\ZN
$$
where  $b=B/N$ is in the range $0\leq b\leq 1$.
By construction, at each  real real root $\l_k$
 $Z_N(\l) N$ takes the half-integer value (cfr. eqs. \LBAE\ and \ZN) $I_k$:
$$
Z_N(\l_k)={ I_k \over N} ~~.
\eqn\IkN
$$

\REF\devegaNPB{H.J. de Vega, Nucl. Phys. {\bf B}, Proc. Suppl. {\bf 18A}
(1990) 229.}

Let us  solve the BA equations  \LBAE\ in the thermodynamic limit.
For the ground state we assume as usual the regular filling
$$
I_{k+1}-I_k=1 ~~.
$$
Then, after introducing the continuous density of roots
$$
\sigma_b(\l_k)=\lim_{N\to\infty} {1\over N(\l_{k+1}-\l_k)} ~~,
$$
in the $N\to\infty$ limit in which the BA
roots are closely spaced, we have from eq. \IkN\ [\devegaNPB]
$$
{d Z_\infty \over d\l}=\sigma_b(\l) ~~.
$$
Thus in the limit $N\to\infty$ eq. \ZN\ reads
$$
\sigma_b(\l)={1\over 2\pi} \Phi'(\l,\gamma/2)+{b\over \pi}
- {1\over 2\pi}\int_{-\pi/2}^{\pi/2} d\mu \> \Phi'(\l-\m,\g) \> \s_b(\m) ~~,
\eqn\densita
$$
with
$\Phi'(\l,\g)=d\Phi(\l,\g) /d\l$. Eq. \densita\ is identical
to the corresponding equation for the six-vertex model, except for the
constant term $b/\pi$ [\devegaNPB]. This linear equation can be solved by
Fourier series. Starting with
$$
\s_b(\l)=\sum_{k=-\infty}^{+\infty} e^{2ik\l} \s_k
$$
$$
\Phi(\l,\g)=2\l+2\sum_{k=1}^\infty {\sin(2 k\l) \over k} e^{-2k\g} ~~,
$$
we find
$$
\s_b(\l)={1+b \over 2\pi} + {1\over 2\pi} \sum_{ {k=-\infty} \atop
{k\not=0} }^{+\infty} {e^{2ik\l} \over \ch(k\g)} ~~.
\eqn\solFourier
$$
This density results to be equal to the  six-vertex ground state
density of roots  plus the constant $b/(2\pi)$. This implies  that the
number of roots
$$
N\int_{-\pi/2}^{\pi/2} d\l \>  \s_b(\l) ={N+B\over 2}
$$
is $B/2$ larger than in the six-vertex model.

The free energy per site at fixed $b$ is defined as
$$
f_b(\t,\g)=-\lim_{N\to\infty} {1\over N} \log \Lambda(\t)
$$
which, by using eq. \final, takes the form
$$
f_b(\t,\g)=-\log\abs{\sh(\g+\t)} - \t b +i\!\!
\int_{-\pi/2}^{\pi/2} d\l\> \Phi(\l-i\t,\g/2)\> \s_b(\l) ~~.
\eqn\fb
$$
Inserting eq. \solFourier\ in \fb\ yields the $b$-independent result
$$
f_b(\t,\g)=f_0(\t,\g)
$$
in the thermodynamic limit. This shows that all the ground states of
the various $B$-dependent  sectors of the Hilbert space are degenerate
in this limit.

The excited states follow by allowing holes in the half-integers
$I_k$:
$$
I_{k+1}-I_k=1+\sum_{h=1}^{N_h} \delta_{ii_h} ~~.
$$
The corresponding root density and eigenvalues are identical to those
of the six-vertex model since they are $B$-independent.

\chapter{The large $\t$ behaviour of the NIS model and of its Bethe
Ansatz solution.}

It is always instructive to investigate the properties of integrable
models in the limit in which the spectral parameter $\t$ takes large
values. In particular the quantum group structure usually emerges in
such a limit [\devegaNPB].  We see from eq.\pesi\ that the leading
contribution for large $\t$ of the various Bolztmann weights is
$e^{d\t}$ with $d=2,\frac 32,1,\frac 12,0$.  This behaviour makes the
diagonalization of the transfer matrix in the large $\t$ limit as
difficult as for finite values of $\t$. This situation differs from
the one of the six-vertex model where $d=1,0$ and the transfer matrix
in the limit $\t\to\infty$ is easily diagonalizable and non-singular.

Clearly, the purely black configurations
$R^{\mn}_\mn(\t)=R^{\n\m}_{\n\m}(\e-\t)$ [see eqs.\pesi] dominate in
the $\t\to+\infty$ limit. More precisely
$$
R^\mn_{\l\s}(\t)\> \ug_{\t\to+\infty}  \> -{e^{-\e}\over 2} \> e^{2\t}
\delta_{\m\l} \delta_{\n\s}(1-\delta_{\m\s}) ~~.
$$
One then gets for the $N$-sites transfer matrix in the same limit
$$
\tau(\t)\ug_{\t\to+\infty} \left(-{e^{-\e}\over 2}\right)^N e^{2N\t}
M_{{\ss 2N}}  \left(1+o(e^{-\t})\right) ~~,
$$
where
$$
M_{\ss 2N}=\sum_{\set\m} \ket{\m_1\m_2\ldots \m_N}
\bra{\m_2\m_3\ldots \m_N\m_1}~, \quad \m_i\not= \m_{i+1}
\eqn\Mshift
$$
is a left-shift operator in the purely black sector with the condition
that all the neighboring indices are different. Notice that the
operator $M_{\ss 2N}$ has a non-zero kernel since
it annihilates all states having either a couple of
equal neighboring black indices or  at least one  red index.
Thus the diagonalization of $M_{\ss 2N}$ does not yield any
information on the asymptotic behaviour at large $\t$ of the states in
its kernel. We then move to
the next to leading
contribution in $\tau(\t)$. It  is obtained by replacing in two neighboring
sites  the weights
$R^{\mn}_\mn(\t)$ with $R^{\m\a}_{\m\a}(\t)$ and $R^{\a\m}_{\a\m}(\t)$
in all  possible ways. From
$$
R^{\m\a}_{\m\a}(\t) = R^{\a\m}_{\a\m}(\t)
\> \ug_{\t\to+\infty}  \> -{e^{-\e}\over 2} \> e^{{3\over 2}\t}
$$
we find
$$
\tau(\t)\ug_{\t\to+\infty} \left(-{e^{-\e}\over 2}\right)^N \left( e^{2N\t}
M_{{\ss 2N}} +  e^{(2N-1)\t}
M_{{\ss 2N-1}} \right) \left(1+o(e^{-\t}) \right) ~~,
\eqn\leadsublead
$$
where the operator $M_{{\ss 2N-1}}$ is again a left-shift operator of
the form \Mshift\ with all black indices but one which is red.
Just like $M_{\ss 2N}$, the operator $M_{\ss 2N-1}$ has a non-zero kernel.

It is useful to match eq. \leadsublead\ with the large $\t$ behaviour
of the eigenvalues \final\ obtained from the BA equations.
We have
$$
\eqalign{
\Lambda(\t)\ug_{\t\to+\infty}&
\left(-{1 \over 2}\right)^N \> e^{(N+B)\t} e^{i\omega -(N-m)\e}
\left[ 1+e^{-2\t}\left(2\sh\e \sum_{j=1}^m e^{-2i\l_j}
-Ne^{2\e}\right)\right]
\cr
&~~+ \left({1 \over 2}\right)^N \> e^{(N-B)\t} e^{-i\omega +(B-m)\e}
\left[ 1-e^{-2\t}\left(2e^{2\e}\sh\e \sum_{j=1}^m e^{-2i\l_j}
+N\right) \right] \cr
&~~ + \hbox{~lower orders in~} e^{\t} ~~.
}
\eqn\Lasym
$$
Comparison between eqs.\Lasym\ and \leadsublead\ leads to the
conclusion that the highest degree term $e^{2N\t}$ in eq. \leadsublead\
corresponds to an eigenvector $\psi_{\ss 2N}$  of $M_{\ss 2N}$ whose
eigenvalue in the large $\t$ limit is given by \Lasym\ with $B=N$.
The eigenvectors of $M_{\ss
2n}$ are  reference states of the form \refstate\ with all black
indices. Moreover
$$
M_{\ss 2N}\psi_{\ss 2N}=\ep_{\ss L}^j \psi_{\ss N}
$$
since $M_{\ss 2N}$ is a shift operator. Thus from eq. \Lasym\
$e^{i\omega}$ must be a $L$-th root of unity. The operator $M_{\ss 2N}$
annihilates all vectors outside the purely black sector of the Hilbert
space. States with only one red index are instead captured by the
operator $M_{\ss 2N-1}$ which, being a shift operator, has eigenstates
of the form \refstate\ and roots of
unity as eigenvalues. Thus also in this case $e^{i\omega}$ is a root
of unity.
%This subleading behaviour  corresponds to setting $B=N-1$ in eq.
%\nmax.
Again there are no BA roots if $B=N-1$.
\FIG\Operators{Contribution of order $e^{(2N-2)\t}$ to the transfer
matrix in the limit $\t\to\infty$.}

In order to find roots of the BA equations we must move to the next to
subleading term, $e^{(2N-2)\t}$, corresponding to $B=N-2$ in the
asymptotic expansion of $\tau(\t)$ for large $\t$.
The transfer matrix in this limit is
$$
\eqalign{
\tau(\t)\ug_{\t\to+\infty} & \left(-{e^{-\e}\over 2}\right)^N
\left[ e^{2N\t} M_{{\ss 2N}}  +  e^{(2N-1)\t} M_{{\ss 2N-1}}
+ e^{(2N-2)\t} M_{{\ss 2N-2}} \right] \cr
&~~+\left(-{e^{2\t-\e}\over 2}\right)^{N-1} \>
\left[ {e^\e\over 2} M_{\ss 2N} +M
%+\sh\e M^{(1)}_{{\ss 2N-2}} + {e^\e\over 2} M^{(2)}_{{\ss 2N-2}}
%+{e^{\e /2} \over 2} M^{(3)}_{{\ss 2N-2}} \right.\cr
%& \left. +{1-e^{-\e /2} \over 2}M^{(4)}_{{\ss 2N-2}}
%+{1\over 2} M^{(5)}_{{\ss 2N-2}}
%+{e^{\e /2} \over 2} M^{(6)}_{{\ss 2N-2}}
\right] . }
$$
The operator $M_{{\ss 2N-2}}$ is a shift operator of the form \Mshift\
involving 2 red indices. Its eigenstates are then of the kind
\refstate\ and its eigenvalues are roots of unity.
The operator $M$ is  given by
$$
M=\sh\e M^{(1)}_{{\ss 2N-2}} + {e^\e\over 2} M^{(2)}_{{\ss 2N-2}}
+{e^{\e /2} \over 2} M^{(3)}_{{\ss 2N-2}}
 +{1-e^{-\e /2} \over 2}M^{(4)}_{{\ss 2N-2}}
+{1\over 2} M^{(5)}_{{\ss 2N-2}}
+{e^{\e /2} \over 2} M^{(6)}_{{\ss 2N-2}}
$$
where the operators
$M^{(i)}_{{\ss 2N-2}}$ ($i=1,\ldots,6$), as  the diagrammatic
representation of  in Fig.\Operators\ shows, involve a couple of
equal neighboring indices. Eigenvalues of the transfer matrix
 with  asymptotic behaviour  $e^{(2N-2)\t}$ not corresponding to
reference states can then be obtained by diagonalizing the operator
$M$.
Obviously (see Fig.\Operators) the reference states \refstate\ belong
to the kernel of $M$ and conversely the eigenstates of $M$ are in the
kernel of $ M_{{\ss 2N}}$, $M_{{\ss 2N-1}}$ and $ M_{{\ss 2N-2}}$.
Let us start by considering the plane waves $\ket{\ep_{\ss N}^j;\m}$
of eq.\planecouple. Calling $\n_L$ and $\n_R$ the (black) indices on
the left and on the right of the couple $\m\m$ in the states
$\ket{\ep_{\ss N}^j;\m}$, the state
$$
\Psi (\ep_{\NN}^j)=e^{\e /2} \sum_{{\m=1 \atop \m\not= \n_L,\n_R}}^t
 \ket{\ep_{\ss N}^j;\m}
+2e^{-\e /2} \sh\e \ket{\ep_{\ss N}^j;\n_R }
+\sum_{\a=1}^r \ket{\ep_{\ss N}^j;\a}
$$
satisfies
$$
M\Psi (\ep_{\NN}^j)=-{e^\e \over 2} \Psi (\ep_{\NN}^j) ~~,
$$
 where
 eq. \secondor\ has been taken into account. Therefore we have
$$
\tau(\t)\>\Psi (\ep_{\NN}^j)\ug_{\t\to+\infty}
\left(- {1\over 2} \right)^N \ep_{\ss N}^j e^{(N-2)\e} \> e^{2(N-1)\t}
\Psi(\ep_{\NN}^j) ~~.
$$
This corresponds to a solution of the BA equations
with $B=N-2$, $m=2$ and $e^{i\omega}=\ep_{\ss N}^j$  which is a root
of unity.

In conclusion, we have verified for $B=N,N-1,N-2$
that $e^{i\omega}$ is a root of unity. We conjecture that this is
the case for all values of $B$.

\chapter{The partition function and the excitation spectrum
in the thermodynamic limit}

\REF\Pearce{P.A. Pearce, Phys. Rev. Let {\bf 58} (1987) 1502.}
\REF\KlumperI{A. Kl\" umper and J. Zittartz, Z. Phys. {\bf B71} (1988)
495; \nextline A. Kl\" umper, A. Schadchneider and J. Zittartz, Z.
Phys. {\bf B76} (1989) 247.}
\REF\KlumperII{A. Kl\" umper, J. Phys. {\bf A}: Mathematical and
General {\bf 23} (1990) 809. }
In this section we collect some results about the determination of the
free energy per site in the thermodynamic limit and make the
comparison between the result obtained by means of
inversion techniques [\Stroganov,\Pearce-\KlumperII] and the
one we get from the BA equations \BAE.

We start by observing that eq.\inversion\ can be rewritten as
$$
\tau(\t+\e)\>\tau(\t)= \rho(\t)^N \1 +o(e^{-N}) ~~,
\eqn\inversionII
$$
where now we have made explicit the fact that the second term on the
RHS is vanishing in the limit $N\to\infty$.
Correspondingly, for the eigenvalues $\Lambda(\t)$ of $\tau(\t)$ one
gets
$$
\lim_{N\to\infty} \Lambda(\t)\Lambda(\e+\t)=\rho(\t)^N ~~.
\eqn\LIM
$$
For the ground state $\Lambda_0(\t)$ we define
$$
L_0(\t)=-{1\over \sh\e}\lim_{N\to\infty}\Lambda_0(\t)^{1/N}
$$
which, because of \propIII\ and \LIM, satisfies ($\Lambda_0(\t)$ is real)
$$
L_0(\t)L_0(-\t)=
{\sh(\e+\t)\sh(\e-\t) \over \sh^2\e}\equiv \varphi(\t)\varphi(-\t) ~~,
\eqn\iter
$$
where property \propII\  has been taken into account and a suitable
function $\varphi(\t)$ has been introduced. The explicit form of
$\varphi(\t)$ has to be chosen in such a way that the following
properties for $L_0(\t)$
are satisfied: {\it i)} analyticity and absence of zeroes
in the physical strip $-\e\leq {\rm Re}\t\leq\e$, {\it ii)}  $i\pi$
periodicity, which follows from eq. \propII\ under the hypothesis
that the ground state belongs to the even $B$ sector as discussed in the
Sec. 3.

Eq.\iter\ can be solved iteratively. Starting  from
$L_0(\t)=\varphi(\t)$ and imposing alternatively crossing invariance \propIII\
and eq.\iter\
we end up with the solution in terms of an infinite product
$$
L_0(\t)=\prod_{k=0}^\infty
{\varphi(2k\e+\t)\over \varphi((2k+1)\e+\t)} \>
{\varphi((2k+1)\e-\t)\over \varphi(2(k+1)\e-\t)} \>
{\varphi(2(k+1)\e)\over \varphi(2k\e)} ~~,
$$
where the last ratio in the product has been introduced so as to make
$L_0(0)=L_0(\e)=1$. The obvious identification is then
$$
\varphi(\t)={\sh(\e+\t) \over \sh\e}
$$
so that, in the thermodinamic limit,
$$
\Lambda_0(\t)^{1/N}=-\sh\e\> \prod_{k=0}^\infty
{\sh((2k+1)\e+\t)\over \sh(2(k+1)\e+\t)}\>
{\sh(2(k+1)\e-\t)\over \sh((2k+3)\e-\t)}\>
{\sh((2k+3)\e)\over \sh((2k+1)\e)} ~~.
\eqn\SOL
$$
This solution has the required analyticity  and periodicity
properties. By taking the logarithm of $\Lambda_0(\t)$ we obtain the
free energy per site $f_{\rm NIS}(\t,\e)$:
$$
f_{\rm \scriptscriptstyle NIS}(\t,\e)=-\log\Lambda_0(\t)^{1/N} ~~.
$$

The expression \SOL\ for the ground eigenvalue could have been
obtained equally well  from the BA equations \BAE\ with $B=0$.
{}From eq.\fb\ we get for the free energy
$$
f_{{\scriptscriptstyle\rm NIS}}(\t,\e)=\t+\sum_{k=1}^\infty {e^{k\e}\over k}
\> {\sh(2k\t)\over \ch(k\e) } - \log|\sh(\e-\t)| ~~.
\eqn\freeenergy
$$
 The sum on the RHS can be recast in the
following form
$$
\eqalign{
\sum_{k=1}^\infty {e^{k\e}\over k}
{\sh(2k\t)\over \ch(k\e) } =&
\sum_{k=1}^\infty (-1)^k \log{ 1-e^{2(\t+k\e)}\over 1-e^{-2(\t-k\e)} } \cr
=& -\log \prod_{k=1}^\infty
{ 1-e^{-2\t+4k\e}\over 1-e^{-2\t+2(2k-1)\e} }
{ 1-e^{2\t+2(2k-1)\e}\over 1-e^{2\t+4k\e} }
}
$$
so that (see also [\KlumperII])
$$
\eqalign{
\Lambda_0(\t)^{1/N}=& e^{-f(\t,\e)} \cr
=& {1\over 2} e^{-\e} \prod_{k=0}^\infty
{ 1-e^{-2\t+4(k+1)\e} \over 1-e^{-2\t+2(2k+3)\e} }
{ 1-e^{2\t+2(2k+1)\e}\over 1-e^{2\t+4(k+1)\e} } }
%\eqn\SOLII
$$
which is completely equivalent to \SOL\ owing to
$$
 \prod_{k=0}^\infty \> e^{2\e} \>
{\sh((2k+3)\e)\over \sh((2k+1)\e)} =-{e^{-\e} \over 2 \sh\e} ~~.
$$
The derivation  above shows  the relation between the free
energies of the NIS and six vertex models is
$$
f_{{\scriptscriptstyle\rm NIS}}(\t,\e)
=f_{{\scriptscriptstyle\rm 6-V}}(-\t,\gamma)~, \qquad \gamma=-\e ~~.
$$

Following ref. [\KlumperI], the inversion method can also be used for
determining all those eigenvalues $\Lambda(\t)$ such that the limit
$$
l(\t)=\lim_{N\to\infty} {\Lambda(\t) \over \Lambda_0(\t)}
\eqn\excit
$$
is finite and non-zero. From the definition \excit\ and eq.\LIM,  $l(\t)$
satisfies the inversion relation
$$
l(\t)l(\e+\t)=1\> \qquad {\rm or} \qquad l(\t)\overline{l(-\t)}=1 ~,
\eqn\excitinv
$$
where in the last equation crossing invariance
has been taken into account. From
crossing invariance and eq. \excitinv\
 it is immediate to show that $l(\t)$, in addition to
being $i\pi$ periodic owing to property \propII, is also $2\e$ periodic:
$$
l(\t)=\overline{ l(\e-\t)} ={1\over l(\t-\e)}
={1\over \overline{l(2\e-\t)}}=l(\t-2\e) ~.
$$
The excitations, just like the ground state,  ought to be analitic in
the physical strip $-\e\leq {\rm Re}\,\t\leq\e$ but zeros are now
allowed.
The typical ansatz is thus written as a product of Jacobi elliptic
functions of modulus $k$ as follows
$$
l(\t)=\prod_{j=1}^n C_j \>\sn( D(\t_j-i\t)) ~,
$$
with $C_j$ and $D$ coefficients to be determined.
In the following we shall assume $n$ to be even, as in the six-vertex model.
The parameters
$\t_j$ specify the location of the zeroes.  Periodicity $2\e$
and $i\pi$ require
$$
D={K'\over\e} \qquad {\rm and} \qquad {K'\over K}=-{2\e\over\pi} ~~.
\eqn\KK
$$
 By making
use of \REF\grads{I. S. Gradshteyn and I. M.
Ryzhik, {\it Table of integrals, series and products}, XX edition,
Academic Press (1980).}[\grads]
$$
\sn(z-iK')\>\sn(z)={1\over k} ~,
\eqn\snprop
$$
 eqs.\excitinv\ are satisfied if
$$
C_j=\pm\sqrt k \>, \qquad {\rm and} \qquad \{ \bar\t_j\}=\{\t_i-i\e \} ~.
$$
The simplest solution, corresponding to elementary excitations, is
given by
$$
\t_j=\a_j+i\e /2 ~, \qquad 1\leq j\leq n
\eqn\elem
$$
where $\a_k$ real numbers (other more general solutions solutions would
describe bound states). For the $\a_k$ given by eq.\elem\ we get
$$
l(\t)=\pm \prod_{j=1}^n \>\sqrt k \>\sn\left({K'\over\e} (\a_j-i\t)
+i{K' \over 2} \right) ~~.
\eqn\excitsol
$$
The energy per site of the ground state is given by
$$
\eqalign{
\epsilon_0=&\left. {d \> f_{\ss NIS}(\t) \over d\t} \right|_{\t=0} \cr
=& \> 2\sum_{k=1}^\infty {e^{k\e}\over \ch(k\e)} -{e^{-\e}\over \sh\e}
{}~~. }
$$
The momenta and energies of the first excitations can be obtained from
the expression of $l(\t)$. We have
$$
\eqalign{
p-p_0=& -i \ln l(0) =\sum_{j=1}^n p_j ~~,\cr
\epsilon -\epsilon_0 =&
-\left. {d \ln l(\t) \over d\t} \right|_{\t=0}=\sum_{j=1}^n \epsilon(p_j)
{}~~,}
\eqn\excitmomen
$$
which give
$$
p_j=-i\sqrt k \> \log \sn\left( {K'\over \e} \a_j +i{K' \over 2} \right)
\eqn\snp
$$
and, with the help of [\grads],
$$
{d \>\sn u\over du}=\sqrt{(1-\sn^2 u)(1-k^2 \sn^2 u)} ~,
$$
one gets
$$
\epsilon(p_j)={K'\over \e}\sqrt{(1-k)^2+4k\sin^2 p_j} ~~.
\eqn\snen
$$
Notice that $p_j$ is real owing to eq. \snprop\ as it must be.
{}From eq. \snen\   the energy gap is obtained by setting $p_j=0$ with
the minimal value $n=2$ in eqs.\excitmomen:
$$
{\rm gap}=2{K'\over \e}(1-k) ~~.
\eqn\eqngap
$$

In summary, the free energy \freeenergy\ and the  excited states
\excitsol\  of
the general NIS model in the thermodynamic limit are completely
equivalent to those of the six-vertex ones. There is one branch of
excitations with non-zero gap, given by eq. \eqngap,
 and obeying the dispersion relation
\snen.
\REF\HdV{H.J. de Vega and F. Woynarovich, Nucl.Phys. {\bf B251} (1985) 439.}

Finite size corrections can be computed here with the method proposed
in [\HdV]. Since the model is gapful, they are exponential in the
size. More precisely they are typically of  order
$$
k^{+N/2} ~~,
$$
where $k<1$ is defined by eq. \KK.

\chapter{The NIS model as a mixed spin $1/2$ and $\infty$-spin model}

In this section we show that the BA equations and
the transfer matrix eigenvalues of the NIS model derived in Section 3
can be identified  with those of a spin system describing a
mixture of spin $1/2$ and spin $s$ particles in the limit in which
$s\to\infty$. In ref. [\babujian] it has been shown that the ground
state of a pure spin $s$ model is formed by a Dirac sea of solutions
of the BA equation arranged in strings
of $2s$ roots (called $2s$-strings). This situation generalizes the usual
situation encountered in spin $1/2$ systems where the sea is formed by
real roots (1-strings).
It is also known [\mixspin]  that  the ground state of a system with
mixed spin 1/2 and $s=1$
is just the superposition of the two seas. Since the form of the
BA equations found in [\mixspin]
is immediately generalizable to an arbitrary value $s$,
we can conclude  that the ground state of a  system with mixed
1/2 and $s$ spin
 is formed by a sea of
$2s$-strings plus a sea of real roots. Namely, we have $M$ real roots
$\l_j$ ($1\leq j\leq M$) and $\n_s$ $2s$-strings:
$$
\l_{\a,l}=\s_\a+i\left(l+\frac 12\right)\g + \varepsilon_{N_{1/2}N_s}
$$
with $1\leq\a\leq\n_s$ and $-s\leq l\leq s-1$. Here the real numbers
 $\s_\a$ are the strings' centers and   $\varepsilon_{N_{1/2}N_s}$ are
quantities vanishingly small  when ${N_{1/2},\, N_s}\to\infty$.

\noindent
The BA equations (with periodic boundary conditions)
in presence of string solutions are then
$$
\eqalign{
\left( {\sin(\l_k+i\g /2) \over \sin(\l_k-i\g /2) } \right)^{N_{1/2}}
\left( {\sin(\l_k+is\g ) \over \sin(\l_k-is\g ) } \right)^{N_s}
= &
-\prod_{ i=1}^M
 {\sin(\l_k-\l_i+i\g) \over \sin(\l_k-\l_i-i\g) }  \cr
&\qquad \times \prod_{\a=1}^{\n_s}\prod_{j=-s}^{s-1}
 {\sin(\l_k-\l_{\a,j}+i\g) \over \sin(\l_k-\l_{\a,j}-i\g) }
}
\eqn\BAEMIXREAL
$$
for a real root $\l_k$ and
$$
\eqalign{
\left( {\sin(\l_{\a,l}+i\g /2) \over \sin(\l_{\a,l}-i\g /2) } \right)
^{N_{1/2}}
\left( {\sin(\l_{\a,l}+is\g ) \over \sin(\l_{\a,l}-is\g ) } \right)^{N_s}
= &
-\prod_{ {i=1 } }^M
 {\sin(\l_{\a,l}-\l_i+i\g) \over \sin(\l_{\a,l}-\l_i-i\g) }  \cr
&\qquad \times \prod_{\b=1}^{\n_s}\prod_{j=-s}^{s-1}
 {\sin(\l_{\a,l}-\l_{\b,j}+i\g) \over \sin(\l_{\a,l}-\l_{\b,j}-i\g) }  }
\eqn\BAEMIXSTRING
$$
for a string solution $\l_{\a,l}$. The integers
$N_{1/2}$ and $N_s$ denote the number of sites where particles with
spin $1/2$ and $s$ respectively sit. Both $N_{1/2}$ and $N_s$ are
assumed to be even integers.

\noindent
The transfer matrix eigenvalue of  this mixed spin system is given by
$$
\L(u,\vec \l)=\L_A(u,\vec\l)+\L_D(u,\vec\l) ~~,
\eqn\MIXVALUECOMP
$$
with
$$
\eqalign{
\L_A(u,\vec\l)=\sh^{N_{1/2}}(\g-u) &
\left( {\sh\left[(s+1/2)\g-u\right] \over \sh [(s+1/2)\g] }
\right)^{N_s}
\prod_{j=1}^M {\sin(\l_j+iu+i\g /2) \over \sin(\l_j+iu-i\g /2)} \cr
&\qquad \times \prod_{\a=1}^{\n_s}
\prod_{l=-s}^{s-1} {\sin(\l_{\a,l}+iu+i\g /2) \over
\sin(\l_{\a,l}+iu-i\g /2)} ~~, }
\eqn\MIXVALUE
$$
where $\vec\l=(\l_1,\ldots,\l_M,\l_{1,-s},\ldots,\l_{\n_s,s-1})$
are solutions of \BAEMIXREAL\ and \BAEMIXSTRING\
and $\L_D(u,\vec\l)=\overline{ \L_A(\g-u,\vec\l)}$.

\noindent
The infinite $s$ limit of eq. \BAEMIXREAL\ gives then
$$
\left( {\sin(\l_k+i\g /2) \over \sin(\l_k-i\g /2) } \right)^{N_{1/2}}
e^{-2i(N_s-2\n_s)\l_k}
=
- \exp\left({4 i\sum_{\a=1}^{\n_s}\sigma_\a}\right) \prod_{ i=1}^M
 {\sin(\l_k-\l_i+i\g) \over \sin(\l_k-\l_i-i\g) }  ~
\eqn\MIXLIM
$$
which coincide with  eqs.\BAE\ if we set $\g=-\e$, $N_{1/2}=N$,
$B=N_s-2\n_s$ and $e^{i\omega}=\exp({-2
i\sum_{\a=1}^{\n_s}\sigma_\a})$.
In the same way   and with the same
identification of parameters the large $s$ limit of eq.\MIXVALUE\ is
$$
\L_A(u,\vec\l)= \sh^{N_{1/2}}(\g-u)
e^{-(N_s-2\n_s)u} \exp\left( -2i \sum_{\a=1}^{\n_s} \s_\a \right)
\prod_{j=1}^M {\sin(\l_j+iu+i\g /2) \over \sin(\l_j+iu-i\g /2)}
$$
which inserted back in \MIXVALUECOMP\
reproduces exactly the eigenvalue \final\ of the   NIS model if $u=-\t$.
\REF\AAVV{M. Gaudin, Phys. Rev. Lett. {\bf 26} No.26 (1971) 1301. \nextline
A. Kirillov et N.Yu. Reshetikhin, J. Phys. {\bf A}: Math. and
Gen. {\bf 20} (1987) 1565 and 1585. \nextline
H. Frahm, N-C. Yu and M. Fowler, Nucl. Phys. {\bf B336}
(1990) 396. }

One can further explore this correspondence and give a proof,
at least in the limit $N\to\infty$, to our conjecture that
the phase factor $e^{i\omega}$ of the NIS model is in fact a root of
unity. To this purpose we multiply all the BA equations \BAEMIXSTRING\
corresponding to a string with center say $\s_\a$ and take the
logarithm [\AAVV]. After some algebra we obtain in the limit
$N_{1/2},N_s\to\infty$
$$
\eqalign{
N_{1/2}\Phi(\s_\a,s\g)+N_s\sum_{k=1}^{2s} \Phi(\s_\a, & (k-\frac 12)\g) =
2\pi I_\a +
\sum_{\b=1}^{\n_s} \Bigg[ \Phi(\s_\a-\s_\b, 2s\g)  \cr
 +2\sum_{k=1}^{2s-1} \Phi(\s_\a-\s_\b,k\g) \Bigg]
&+\sum_{j=1}^{M} \left[ \Phi\left(\s_\a-\l_j,\left(s+\frac 12\right)\g\right)
+\Phi\left(\s_\a-\l_j,\left(s-\frac 12\right)\g\right)  \right] ~~, }
\eqn\REAL
$$
where, for real $\l$,
$$
\Phi(\l,\g)=2\tan^{-1}[ \tan\l \> {\rm coth} \g] .
$$
 As usual $I_\a$ denotes a half-integer.
 Eqs. \REAL\ are  a set of equations for the real roots and
the centers of the strings. We can easily solve for the latter in the
infinite $s$ limit in which \REAL\ reduces to
$$
(N_{1/2}+4sN_s)\s_\a=2\pi I_\a
+8s\sum_{\b=1}^{\n_s}(\s_\a-\s_\b) +2\sum_{j=1}^{M}(\s_\a-\l_j)
+o(s^0) ~~.
$$
Under the hypothesis of regular filling for the ground state
$$
M= {N_{1/2} \over 2}+o(1)~, \qquad \sum_{j=1}^{M}\l_j =o(1)
$$
we get
$$
\s_\a={\pi\over 2sB} \left(I_\a-{2\over
N_{s}}\sum_{\b=1}^{\n_s} I_\b \right) ~~.
$$
This result shows that  the quantity
$$
\exp\left(2i\sum_{\b=1}^{\n_s}\s_\a\right) =\exp \left({ i\pi \over sN_s
}\sum_{\b=1}^{\n_s}I_\a \right)
$$
appearing in \MIXLIM\ is indeed a root of unity and the correspondence with
the eigenvalues of the NIS model holds.

Recall that the eigenvalue of $\hat S_z$ on a state determined by eqs.
\BAEMIXREAL-\MIXVALUE\ is given by
$$
S_z=s(N_s-2\n_s)+\left({N_{1/2}\over 2}-M\right) ~~.
$$
The first term is the contribution of the $s$ spins and the second one
of the spins 1/2. Then if we define
$$
\hat B=\lim_{s\to\infty} {\hat S_z \over s} ~~.
$$
Hence $B$ can be interpreted as a properly normalized eigenvalue of
$\hat S_z$ for $s=\infty$.

\bigskip

\ack
\noindent
GG was supported by the Commission of the European Communities
through contract No. SC900376.

\refout
\figout

\end